\documentclass[prd,tightenlines]{revtex4}
\usepackage{epsfig}
\usepackage{graphics}
\usepackage{latexsym}
\usepackage{amsmath}
\usepackage{amssymb}
\usepackage{rotating}

\begin{document}

ADP-12-40/T807

\bigskip

\title{Chiral extrapolation of nucleon magnetic moments at
next-to-leading-order}

\author{P. Wang$^{ab}$}
\author{D. B. Leinweber$^c$}
\author{A. W. Thomas$^{cd}$}
\author{R. D. Young$^{cd}$}

\affiliation{$^a$Institute of High Energy Physics, CAS, P. O. Box
918(4), Beijing 100049, China}

\affiliation{$^b$Theoretical Physics Center for Science Facilities,
CAS, Beijing 100049, China}

\affiliation{ $^c$Special Research Center for the Subatomic Structure
  of Matter (CSSM), School of Chemistry \& Physics, University of
  Adelaide, SA 5005, Australia}

\affiliation{ $^d$ ARC Centre of Excellence in Particle Physics
at the Terascale,
School of Chemistry \& Physics, University of
  Adelaide, SA 5005, Australia}

\begin{abstract}
Nucleon magnetic moments display a rich nonanalytic dependence on the
quark mass in both quenched and full QCD.  They provide a forum for a
detailed examination of the connection between quenched and full QCD
made possible through the formalism of finite-range regularised chiral
effective field theory.  By defining meson-cloud and core
contributions through the careful selection of a regularisation scale,
one can correct the meson cloud of quenched QCD to make full QCD
predictions.  Whereas past success is based on unquenching the
leading-order loop contributions, here we extend and test the
formalism including next to leading-order (NLO) loop contributions.
We discuss the subtleties associated with working at NLO and
illustrate the role of higher-order corrections.
\end{abstract}


\maketitle

\smallskip

\section{Introduction}

The study of the properties of hadrons continues to attract
significant interest in the process of revealing and understanding the
essential mechanisms of QCD, the fundamental theory of the strong
interactions.  The non-perturbative properties of QCD and the
difficulty of numerically simulating the theory at the light quark
masses of Nature has provided a rich history of phenomenological
models at both the quark and hadronic levels and an intense study of
the quark-mass dependence of hadronic observables in effective field
theory (EFT).

Based on the observation that {\it all} hadron properties show a slow,
smooth variation with quark mass for pion masses above $m_\pi \sim
0.4$ GeV, one can conclude that the nonanalytic contributions from
pion loops are suppressed there~\cite{Thomas:2002sj}.  An alternative
regularization method, namely finite-range-regularization (FRR),
resums the chiral expansion in a manner that suppresses loop
contributions at large pion masses.  Inspired by quark models
\cite{Lu,Lyubovitskij,Faessler} that account for the finite-size of
the nucleon as the source of the pion cloud, FRR EFT has been used to
describe lattice data over a wide range of pion masses.

FRR EFT was first applied in the extrapolation of the nucleon mass and
magnetic moments~\cite{Leinweber:1998ej,Leinweber0,Leinweber1}. The
remarkably improved convergence properties of the FRR expansion mean
that lattice data at large pion masses can be described very well and
the nucleon mass obtained at the physical pion mass compared favorably
with the experimental value.  Later, the FRR method was applied to
extrapolate the vector meson mass, magnetic moments, magnetic form
factors, strange form factors, charge radii, first moments of GPDs,
etc.\ \cite{%
Alton,Armour,Young,Wang1,Leinweber:2004tc,Leinweber:2006ug,Wang2,Wang3,Wang4}.
The results are reasonable and reflect the manner in which FRR EFT
characterizes the essential features of QCD at the hadronic level.

The prevalence of the quenched approximation in the history of lattice
QCD simulations provided an opportunity to explore the possible
connection between quenched and full QCD data.  Indeed quenched chiral
perturbation theory was developed
\cite{Labrenz:1996jy,Savage:2001dy,Leinweber:2002qb,Arndt:2003ww,Tiburzi:2004mv}
to understand how the nonanalytic structure of quenched QCD differed
from that of full QCD.  It was the advent of FRR EFT that made it
possible to define a pion-cloud contribution to hadronic observables
and then proceed to correct the quenched cloud to
that of full QCD~\cite{Young:2002cj}.

The meson loop contributions are calculated in both quenched and
full QCD in terms of the axial coupling constants. One then fits the
quenched FRR EFT to lattice QCD to learn the low energy coefficients.
This is done by fitting the coefficients of the residual series of
terms analytic in the quark mass.  With the assumption that the SU(3) axial
coupling constants, $F, \, D$ and ${\cal C}$, do not differ
significantly between quenched and
full QCD, one can replace the quenched meson cloud contribution of FRR
EFT with the full QCD cloud contribution.  We note that in dimensional
regularization for example, the low-energy coefficients are composed
of both residual series and loop contributions with no recourse to
separating the origin of terms contributing to the total renormalised
coefficient.  On the contrary, in FRR the low energy coefficients of
the residual analytic expansion provide the core contribution which is
considered invariant in moving from quenched QCD to full QCD.

Quenched and partially-quenched FRR chiral EFT has been used to study
baryon electromagnetic phenomena including charge radii, strange
magnetic moments and strange form factors
\cite{Leinweber:2004tc,Leinweber:2006ug,Wang2,Wang3}.  In the previous
calculations, only the leading-order diagrams were included and
unquenched.  For example, power counting with $M_\Delta = M_N$ for the
magnetic form factor, only the leading nonanalytic terms proportional
to $\log(m_\pi)$ and $m_\pi$ were included.  In this paper, we will
include the next-to-leading-order (NLO) contributions.

The paper is organized in the following way. In section II, we briefly
introduce the relevant chiral Lagrangian.  In section III, we study
the nucleon magnetic moments using chiral perturbation theory with FRR
at NLO.  Numerical results and discussions are
presented in section IV.  Finally, section V provides a summary.

\section{Chiral Lagrangian}

There are many papers which deal with heavy baryon chiral
perturbation theory -- for details see, for example, Refs.
\cite{Jenkins:1990jv,Bernard:1992qa,Bernard:2007zu}. For
completeness, we briefly introduce the formalism in this section. In
heavy-baryon chiral perturbation theory, the lowest chiral
Lagrangian for the baryon-meson interaction which will be used in
the calculation of the nucleon magnetic moments, including the octet
and decuplet baryons, is expressed as
\begin{eqnarray}
{\cal L}_v &=&i{\rm Tr}\bar{B}_v(v\cdot {\cal D}) B_v+2D{\rm
Tr}\bar{B}_v S_v^\mu\{A_\mu,B_v\} +2F{\rm Tr}\bar{B}_v
S_v^\mu[A_\mu,B_v]
\nonumber \\
&& -i\bar{T}_v^\mu(v\cdot {\cal D})T_{v\mu} +{\cal C}(\bar{T}_v^\mu
A_\mu B_v+\bar{B}_v A_\mu T_v^\mu),
\end{eqnarray}
where $S_\mu$ is the covariant spin-operator defined as
\begin{equation}
S_v^\mu=\frac i2\gamma^5\sigma^{\mu\nu}v_\nu.
\end{equation}
Here, $v^\nu$ is the nucleon four velocity (in the rest frame, we
have $v^\nu=(1,0)$). D, F and $\cal C$ are the axial coupling constants.
The chiral covariant derivative $D_\mu$ is written as $D_\mu
B_v=\partial_\mu B_v+[V_\mu,B_v]$. The pseudoscalar meson octet
couples to the baryon field through the vector and axial vector
combinations
\begin{equation}
V_\mu=\frac12(\zeta\partial_\mu\zeta^\dag+\zeta^\dag\partial_\mu\zeta),~~~~~~
A_\mu=\frac12(\zeta\partial_\mu\zeta^\dag-\zeta^\dag\partial_\mu\zeta),
\end{equation}
where
\begin{equation}
\zeta=e^{i\phi/f}, ~~~~~~ f=93~{\rm MeV}.
\end{equation}
The matrix of pseudoscalar fields $\phi$ is expressed as
\begin{eqnarray}
\phi=\frac1{\sqrt{2}}\left(
\begin{array}{lcr}
\frac1{\sqrt{2}}\pi^0+\frac1{\sqrt{6}}\eta & \pi^+ & K^+ \\
\pi^- & -\frac1{\sqrt{2}}\pi^0+\frac1{\sqrt{6}}\eta & K^0 \\
K^- & \bar{K}^0 & -\frac2{\sqrt{6}}\eta
\end{array}
\right).
\end{eqnarray}
$B_v$ and $T^\mu_v$ are the velocity dependent new fields which are
related to the original baryon octet and decuplet fields $B$ and
$T^\mu$ by
\begin{equation}
B_v(x)=e^{im_N \not v v_\mu x^\mu} B(x),
\end{equation}
\begin{equation}
T^\mu_v(x)=e^{im_N \not v v_\mu x^\mu} T^\mu(x).
\end{equation}
In the chiral $SU(3)$ limit, the octet baryons will have the same
mass $m_B$. In our calculation, we use the physical masses for the
baryon octets and decuplets. The explicit form of the baryon octet
is written as
\begin{eqnarray}
B=\left(
\begin{array}{lcr}
\frac1{\sqrt{2}}\Sigma^0+\frac1{\sqrt{6}}\Lambda &
\Sigma^+ & p \\
\Sigma^- & -\frac1{\sqrt{2}}\Sigma^0+\frac1{\sqrt{6}}\Lambda & n \\
\Xi^- & \Xi^0 & -\frac2{\sqrt{6}}\Lambda
\end{array}
\right).
\end{eqnarray}
For the baryon decuplet, the symmetric tensor carries three indices
and is defined as
\begin{eqnarray}
T_{111}=\Delta^{++}, ~~ T_{112}=\frac1{\sqrt{3}}\Delta^+, ~~
T_{122}=\frac1{\sqrt{3}}\Delta^0, \\ \nonumber T_{222}=\Delta^-, ~~
T_{113}=\frac1{\sqrt{3}}\Sigma^{\ast,+}, ~~
T_{123}=\frac1{\sqrt{6}}\Sigma^{\ast,0}, \\ \nonumber
T_{223}=\frac1{\sqrt{3}}\Sigma^{\ast,-}, ~~
T_{133}=\frac1{\sqrt{3}}\Xi^{\ast,0}, ~~
T_{233}=\frac1{\sqrt{3}}\Xi^{\ast,-}, ~~ T_{333}=\Omega^{-}.
\end{eqnarray}

The octet, decuplet and octet-decuplet transition magnetic moment
operators are needed in the one-loop calculation of nucleon magnetic
form factors. The baryon octet magnetic Lagrangian is written as:
\begin{equation}\label{lomag}
{\cal L}=\frac{e}{4m_N}\left(\mu_D{\rm Tr}\bar{B}_v \sigma^{\mu\nu}
\left\{F^+_{\mu\nu},B_v\right\}+\mu_F{\rm Tr}\bar{B}_v
\sigma^{\mu\nu} \left[F^+_{\mu\nu},B_v \right]\right),
\end{equation}
where
\begin{equation}
F^+_{\mu\nu}=\frac12\left(\zeta^\dag F_{\mu\nu}Q\zeta+\zeta
F_{\mu\nu}Q\zeta^\dag\right).
\end{equation}
$Q$ is the charge matrix $Q=$diag$\{2/3,-1/3,-1/3\}$. At the lowest
order, the Lagrangian will generate the following nucleon magnetic
moments:
\begin{equation}\label{treemag}
\mu_{p}^{\text{tree}}=\frac13\mu_D+\mu_F,~~~~~~ \mu_{n}^{\text{tree}}=-\frac23\mu_D.
\end{equation}

The decuplet magnetic moment operator is expressed as
\begin{equation}
{\cal L}=-i\frac{e}{m_N}\mu_C
q_{ijk}\bar{T}^\mu_{v,ikl}T^\nu_{v,jkl} F_{\mu\nu},
\end{equation}
where $q_{ijk}$ and $q_{ijk}\mu_C$ are the charge and magnetic
moment of the decuplet baryon $T_{ijk}$. The transition magnetic
operator is
\begin{equation}
{\cal L}=i\frac{e}{2m_N}\mu_T F_{\mu\nu}\left(\epsilon_{ijk}Q^i_l
\bar{B}^j_{vm} S^\mu_v T^{\nu,klm}_v+\epsilon^{ijk}Q^l_i
\bar{T}^\mu_{v,klm} S^\nu_v B^m_{vj}\right).
\end{equation}
In Ref.~\cite{Durand1}, the authors used $\mu_u$, $\mu_d$ and
$\mu_s$ instead of the $\mu_C$ and $\mu_T$. For the particular
choice, $\mu_s=\mu_d=-\frac12 \mu_u$, one finds the following
relationship:
\begin{equation}
\mu_D=\frac32 \mu_u, ~~~ \mu_F=\frac23 \mu_D, ~~~ \mu_C=\mu_D, ~~~
\mu_T=-4\mu_D.
\end{equation}
In our numerical calculations, the above formulas are used and
therefore all baryon magnetic moments are related to one parameter,
$\mu_D$.

In the heavy-baryon formalism, the propagators of the octet or
decuplet baryon, $j$, are expressed as
\begin{equation}
\frac i {v\cdot k-\delta^{jN}+i\varepsilon} ~~{\rm and}~~ \frac
{iP^{\mu\nu}} {v\cdot k-\delta^{jN}+i\varepsilon},
\end{equation}
where $P^{\mu\nu}$ is $v^\mu v^\nu-g^{\mu\nu}-(4/3)S_v^\mu S_v^\nu$.
$\delta^{ab}=m_b-m_a$ is the mass difference of between the two
baryons.  The propagator of meson $j$ ($j=\pi$, $K$, $\eta$) is the
usual free propagator, {\it i.e.}
\begin{equation}
\frac i {k^2-M_j^2+i\varepsilon}.
\end{equation}

\section{Nucleon Magnetic Moments}

In the heavy baryon formalism, the nucleon form factors are defined
as:
\begin{equation}
<B(p^\prime)|J_\mu|B(p)>=\bar{u}(p^\prime)\left\{v_\mu
G_E(Q^2)+\frac{i\epsilon_{\mu\nu\alpha\beta}v^\alpha S_v^\beta
q^\nu}{m_N}G_M(Q^2)\right\}u(p),
\end{equation}
where $q=p^\prime-p$ and $Q^2=-q^2$. According to the Lagrangian, the
one-loop Feynman diagrams which contribute to the nucleon magnetic
moments are plotted in Fig.~1. The intermediate baryons can be octets
and decuplets. Diagrams (a) and (b) are for the leading order, while
diagrams (c), (d), (e) and (f) enter at NLO. The
last two diagrams exist only in the quenched case where
the $\eta'$ is degenerate with the pion and no $K$-meson
loops contribute.

\begin{figure}[tbp]
\begin{center}
\includegraphics[scale=0.45]{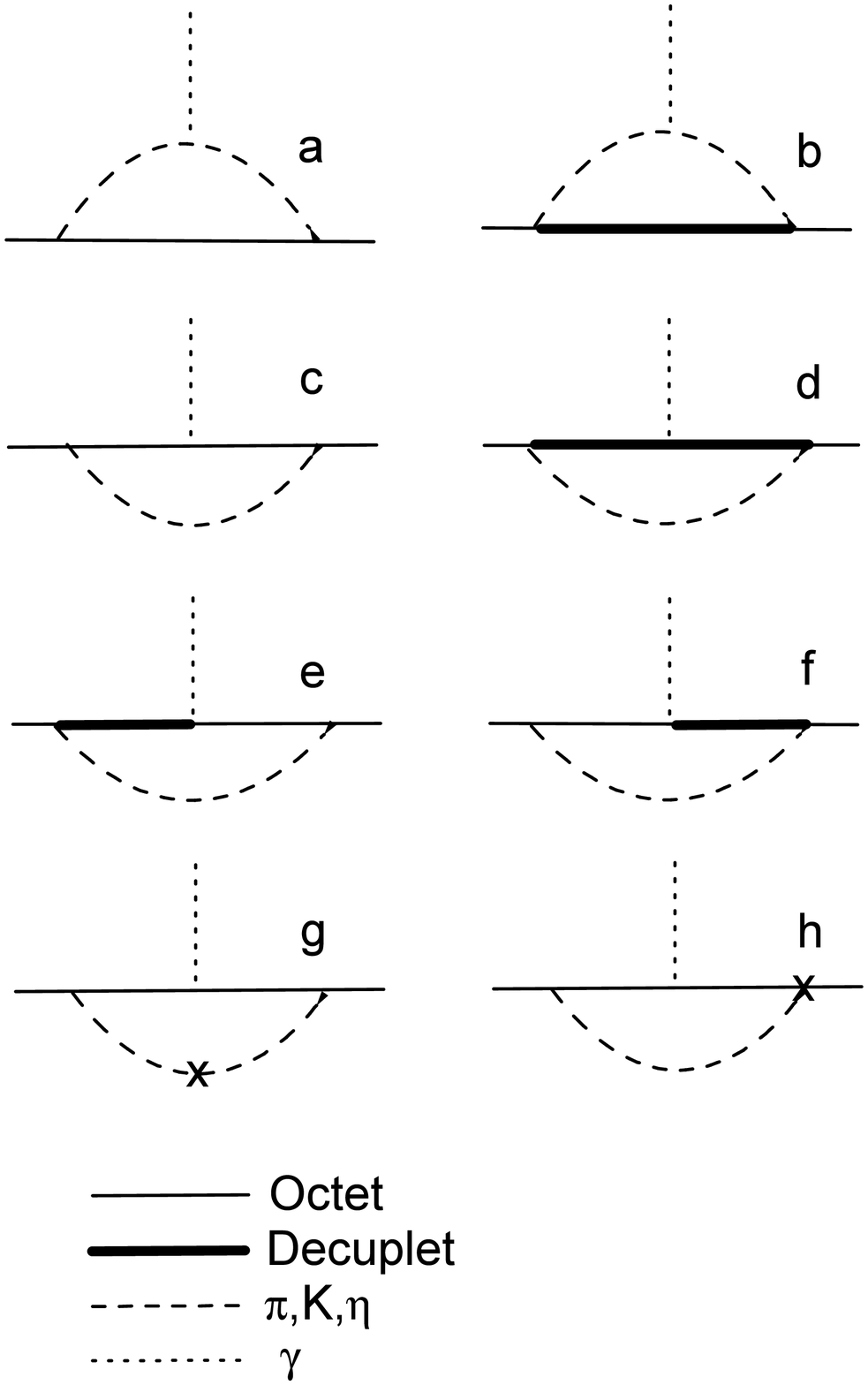}
\caption{Feynman diagrams for the nucleon magnetic moments. The last
two diagrams, (g) and (h), only exist in the quenched case.}
\label{diagrams}
\end{center}
\end{figure}

The loop contribution to nucleon magnetic form factors at leading
order is expressed as
\begin{equation}\label{p1a}
G_M^{p(\text
{LO})}=\frac{m_N}{8\pi^3f_\pi^2}\left[\beta_{1\pi(p)}^{NN}I_{1\pi}^{NN}
+\beta_{1K(p)}^{N\Lambda}I_{1K}^{N\Lambda}+\beta_{1K(p)}^{N\Sigma}I_{1K}^{N\Sigma}
+\beta_{1\pi(p)}^{N\Delta}I_{1\pi}^{N\Delta}+\beta_{1K(p)}^{N\Sigma^\ast}I_{1K}^{N\Sigma^\ast}
\right],
\end{equation}
\begin{equation}\label{n1a}
G_M^{n(\text
{LO})}=\frac{m_N}{8\pi^3f_\pi^2}\left[\beta_{1\pi(n)}^{NN}I_{1\pi}^{NN}
+\beta_{1K(n)}^{N\Sigma}I_{1K}^{N\Sigma}
+\beta_{1\pi(n)}^{N\Delta}I_{1\pi}^{N\Delta}+\beta_{1K(n)}^{N\Sigma^\ast}I_{1K}^{N\Sigma^\ast}
\right].
\end{equation}
The integration $I_{1j}^{\alpha\beta}$ is expressed as
\begin{equation}
I_{1j}^{\alpha\beta}=\int d^3k\frac{k_y^2
u(\overrightarrow{k}+\overrightarrow{q}/2)
u(\overrightarrow{k}-\overrightarrow{q}/2)(\omega_j(\overrightarrow{k}+\overrightarrow{q}/2)
+\omega_j(\overrightarrow{k}-\overrightarrow{q}/2)+\delta^{\alpha\beta})}
{A_j^{\alpha\beta}},
\end{equation}
where
\begin{eqnarray}
A_j^{\alpha\beta}&=&\omega_j(\overrightarrow{k}+\overrightarrow{q}/2)
\omega_j(\overrightarrow{k}-\overrightarrow{q}/2)
(\omega_j(\overrightarrow{k}+\overrightarrow{q}/2)+\delta^{\alpha\beta})
\nonumber \\
&&(\omega_j(\overrightarrow{k}-\overrightarrow{q}/2)+\delta^{\alpha\beta})
(\omega_j(\overrightarrow{k}+\overrightarrow{q}/2)+\omega_j(\overrightarrow{k}-\overrightarrow{q}/2)).
\end{eqnarray}
$\omega_j(\overrightarrow{k})=\sqrt{m_j^2+\overrightarrow{k}^2}$ is
the energy of the meson $j$. $\delta^{\alpha\beta}$ is the mass
difference between baryon $\alpha$ and $\beta$. In our calculation
we use finite-range regularization with $u(\overrightarrow{k})$
the ultra-violet regulator. This leading order contribution has
been studied in the previous paper which gives the leading analytic
term to the magnetic moments.  The first terms in Eqs.\ (\ref{p1a})
and (\ref{n1a}) come from the $\pi$ meson cloud, while the
last two terms correspond to the case where the intermediate baryons
are decuplets.

The NLO contribution to the form factors is expressed
as
\begin{eqnarray}\nonumber
G_M^{p(\text{NLO})}&=&\frac{1}{48\pi^3f_\pi^2}\left[\beta_{2\pi(p)}^{NN}I_{2\pi}^{NN}
+\beta_{2K(p)}^{N\Sigma}I_{2K}^{N\Sigma}+\beta_{2K(p)}^{N\Lambda}I_{2K}^{N\Lambda}
+\beta_{5K(p)}^{N\Lambda\Sigma}I_{5K}^{N\Lambda\Sigma}
+\beta_{2\eta(p)}^{NN}I_{2\eta}^{NN}+\beta_{2\pi(p)}^{N\Delta}I_{2\pi}^{N\Delta}+\beta_{2K(p)}^{N\Sigma^\ast}I_{2K}^{N\Sigma^\ast}
\right. \\
&& \left. +
\beta_{3\pi(p)}^{N\Delta}I_{3\pi}^{N\Delta}+\beta_{5K(p)}^{N\Sigma\Sigma^\ast}I_{5K}^{N\Sigma\Sigma^\ast}
+\beta_{5K(p)}^{N\Lambda\Sigma^\ast}I_{5K}^{N\Lambda\Sigma^\ast}\right],
\end{eqnarray}
\begin{eqnarray}\nonumber
G_M^{n(\text{NLO})}&=&\frac{1}{48\pi^3f_\pi^2}\left[\beta_{2\pi(n)}^{NN}I_{2\pi}^{NN}
+\beta_{2K(n)}^{N\Sigma}I_{2K}^{N\Sigma}+\beta_{2K(n)}^{N\Lambda}I_{2K}^{N\Lambda}
+\beta_{5K(n)}^{N\Lambda\Sigma}I_{5K}^{N\Lambda\Sigma}
+\beta_{2\eta(n)}^{NN}I_{2\eta}^{NN}+\beta_{2\pi(n)}^{N\Delta}I_{2\pi}^{N\Delta}+\beta_{2K(n)}^{N\Sigma^\ast}I_{2K}^{N\Sigma^\ast}
\right. \\
&& \left. +
\beta_{3\pi(n)}^{N\Delta}I_{3\pi}^{N\Delta}+\beta_{5K(n)}^{N\Sigma\Sigma^\ast}I_{5K}^{N\Sigma\Sigma^\ast}
+\beta_{5K(n)}^{N\Lambda\Sigma^\ast}I_{5K}^{N\Lambda\Sigma^\ast}\right],
\end{eqnarray}
where
\begin{equation}
I_{2j}^{\alpha\beta}=\int d^3k\frac{k^2 u^2(\overrightarrow{k})}
{\omega_j(\overrightarrow{k})(\omega_j(\overrightarrow{k})+\delta^{\alpha\beta})^2},
\end{equation}
\begin{equation}
I_{5j}^{\alpha\beta\gamma}=\int d^3k\frac{k^2
u^2(\overrightarrow{k})}
{\omega_j(\overrightarrow{k})(\omega_j(\overrightarrow{k})+\delta^{\alpha\beta})
(\omega_j(\overrightarrow{k})+\delta^{\alpha\gamma}))},
\end{equation}
\begin{equation}
I_{3j}^{\alpha\beta}=\int d^3k\frac{k^2 u^2(\overrightarrow{k})}
{\omega_j(\overrightarrow{k})^2(\omega_j(\overrightarrow{k})+\delta^{\alpha\beta})}.
\end{equation}

All the coefficients $\beta$ in front of the integrals
are shown in Table I for full QCD. The coefficients of the leading-order
contributions are functions of the coupling constants $D$, $F$ and
${\cal C}$.  The coefficients of the NLO contribution are
associated with the tree level baryon magnetic moments.

\begin{table}
\caption{Coefficients $\beta$ for the magnetic moments in full QCD
case.}
\begin{center}
\begin{tabular}{||c|c|c|c|c|c||}
\hline
& $\beta_{1\pi}^{NN}$ & $\beta_{1K}^{N\Lambda}$ &
$\beta_{1K}^{N\Sigma}$ & $\beta_{1\pi}^{N\Delta}$ & $\beta_{1K}^{N\Sigma^*}$ \\
\hline Proton & $(D+F)^2$ & $\frac{(D+3F)^2}{6}$ &
$\frac{(D-F)^2}{2}$ & $\frac{2{\cal C}^2}{9}$ & $-\frac{{\cal
C}^2}{18}$ \\
Neutron & $-(D+F)^2$ & $-$ & $(D-F)^2$ & $-\frac{2{\cal C}^2}{9}$ &
$-\frac{{\cal C}^2}{9}$ \\
\hline
\noalign{\bigskip}
\hline
& $\beta_{2\pi}^{NN}$ & $\beta_{2K}^{N\Lambda}$ &
$\beta_{2K}^{N\Sigma}$ & $\beta_{2\pi}^{N\Delta}$ & $\beta_{2K}^{N\Sigma^*}$ \\
\hline
Proton & $\frac{(D+F)^2}{4}(\mu_D-\mu_F)$ &
$\frac{(D+3F)^2}{12}\mu_D$ & $-\frac{(D-F)^2}{4}(\mu_D+2\mu_F)$ &
$\frac{40{\cal C}^2}{27}\mu_C$ & $\frac{5{\cal C}^2}{27}\mu_C$ \\
Neutron & -$\frac{(D+F)^2}{2}\mu_F$ & $\frac{(D+3F)^2}{12}\mu_D$ &
$-\frac{(D-F)^2}{4}(\mu_D-2\mu_F)$ & $-\frac{10{\cal C}^2}{27}\mu_C$ & $-\frac{5{\cal C}^2}{27}\mu_C$ \\
\hline
\noalign{\bigskip}
\hline
& $\beta_{2\eta}^{NN}$ & $\beta_{3\pi}^{N\Delta}$ &
$\beta_{5K}^{N\Lambda\Sigma}$ & $\beta_{5K}^{N\Sigma\Sigma^*}$ &
$\beta_{5K}^{N\Lambda\Sigma^*}$ \\
\hline
Proton & $-\frac{(D-3F)^2}{12}(\mu_D+3\mu_F)$ &
$\frac{4(D+F){\cal C}}{9}\mu_T$ & $\frac{(D-F)(D+3F)}{6}$ &
$\frac{5(D-F){\cal C}}{18}$ & $\frac{(D+3F){\cal C}}{18}$ \\
Neutron & $-\frac{(D-3F)^2}{6}\mu_D$ & $-\frac{4(D+F){\cal
C}}{9}\mu_T$ & $-\frac{(D-F)(D+3F)}{6}$ & $\frac{(D-F){\cal C}}{18}$
& -$\frac{(D+3F){\cal C}}{18}$ \\
\hline
\end{tabular}
\end{center}
\end{table}

The magnetic moment is defined as $\mu=G_M(Q^2=0)$. The total
nucleon magnetic moments can be written as
\begin{equation}\label{mup}
\mu_p(m_\pi^2)=a^p_0+a^p_2 m^2_\pi+a^p_4 m^4_\pi+(Z-1)\mu_p^{\text{tree}}
+G_M^{p(\text{LO})}(Q^2=0)+G_M^{p(\text{NLO})}(Q^2=0),
\end{equation}
\begin{equation}\label{mun}
\mu_n(m_\pi^2)=a^n_0+a^n_2 m^2_\pi+a^n_4 m^4_\pi+(Z-1)\mu_n^{\text{tree}}
+G_M^{n(\text{LO})}(Q^2=0)+G_M^{n(\text{NLO})}(Q^2=0),
\end{equation}
where the wave function renormalization can be calculated as
\begin{equation}
Z=1-\frac{1}{48\pi^3f_\pi^2}\left[\beta_{\pi}^{NN}I_{2j}^{NN}-\beta_{\pi}^{N\Delta}I_{2j}^{N\Delta}
-\beta_{K}^{N\Lambda}I_{2j}^{N\Lambda}
-\beta_{K}^{N\Sigma}I_{2j}^{N\Sigma}-\beta_{K}^{NN}I_{2j}^{N\Sigma^*}-\beta_{\eta}^{NN}I_{2j}^{NN}\right].
\end{equation}
The coefficients $\beta$ in the wave function renormalization are
listed in Table II.

With the exception of Figs.~\ref{diagrams} (a) and (b), the
contributions of the diagrams in Fig.~1 are proportional to the
tree-level moments, $\mu_{p(n)}^{\text{tree}}$ expressed in
Eq.(\ref{treemag}).
%
% Because we need to determine the fitting
% parameters $a_i$ instead of renormalized low energy constant $\mu_D$,
% it is convenient to replace $\mu_{p(n)}^{\text{tree}}$ by
% $\mu_{p(n)}$.
%
In the quenched case \cite{Young}, the logarithmic divergence of the
magnetic moment encountered in the chiral limit makes it necessary to
replace the leading order estimate $\mu_{p(n)}^{\text{tree}}$ with the
renormalised moment, effectively incorporating physics associated with
higher-order terms of the expansion.  To provide a connection between
the quenched and full QCD expansions, we make this replacement for the
full QCD case as well.  Therefore, the expression for nucleon magnetic
moments can be written as
\begin{equation}
\mu_{p(n)} = a_0^{p(n)}+a_2^{p(n)}m_\pi^2+a_4^{p(n)}m_\pi^4 + \mu_{l1}^{p(n)} + (Z-1)\mu_{p(n)}+\frac{\mu_{l2}^{p(n)}}{\mu_{p(n)}^{\text{tree}}}\mu_{p(n)},
\end{equation}
where $\mu_{l1}^{p(n)}$ is the loop contribution from diagrams (a) and (b) in Fig.~1, while $\mu_{l2}^{p(n)}$ is the contribution from (c), (d), (e) and (f)
expressed in the previous formulas. The above formula can be rewritten as
\begin{equation}\label{renor}
\mu_{p(n)} = \left\{a_0^{p(n)}+a_2^{p(n)}m_\pi^2+a_4^{p(n)}m_\pi^4 + \mu_{l1}^{p(n)}\right\} /(2-Z-\frac{\mu_{l2}^{p(n)}}{\mu_{p(n)}^{\text{tree}}}).
\end{equation}

\begin{table}
\caption{Coefficients $\beta$ and $\tilde{\beta}$ for the wave
function renormalization in full QCD and quenched case.}
\begin{center}
\begin{tabular}{||c|c|c|c|c|c|c||}
\hline
Full QCD & $\beta_{\pi}^{NN}$ & $\beta_{\pi}^{N\Delta}$ &
$\beta_{K}^{N\Lambda}$ & $\beta_{K}^{N\Sigma}$ &
$\beta_{K}^{N\Sigma^*}$ & $\beta_\eta^{NN}$ \\
\hline  & $\frac{9}{4}(D+F)^2$ & $2{\cal C}^2$ &
$\frac{1}{4}(3F+D)^2$ & $\frac{15}{4}(D-F)^2$ & $\frac{5}{6}{\cal C}^2$ & $\frac{1}{2}(3F-D)^2$ \\
\hline
\noalign{\bigskip}
\hline
Quenched & $\tilde{\beta}_{\pi}^{NN}$ &
$\tilde{\beta}_{\pi}^{N\Delta}$ & $\tilde{\beta}_{\eta}^{NN}$ &
$\tilde{\beta}_{dh}^{NN}$ & $\tilde{\beta}_{h}^{NN}$ & \\
\hline & $-\frac{9}{4}D^2-\frac{9}{4}F^2+\frac{15}{2}DF$ &
$\frac{1}{2}{\cal C}^2$ & $-\frac{3}{2}D^2-\frac{3}{2}F^2+DF$ &
$\frac{3}{4}M_0^2(3F-D)^2$
& $3(3F-D)(D-F)$ & \\
\hline
\end{tabular}
\end{center}
\end{table}

Since the lattice data of the magnetic moment are obtained in the quenched
approximation, we should fit the lattice data using quenched chiral
perturbation theory. In the quenched case, only the pion loop makes a
contribution. The coefficients in the quenched case are shown in table
III. They can be obtained following the methodology of
Ref.~\cite{Leinweber:2002qb}. Remember, in this case, we have two more
diagrams, i.e. (g) and (h) in Fig.~1.

\begin{table}
\caption{Coefficients $\tilde{\beta}$ for the magnetic moments in
quenched case.}
\begin{center}
\begin{tabular}{||c|c|c|c|c||}
\hline
& $\tilde{\beta}_{1\pi}^{NN}$ & $\tilde{\beta}_{1\pi}^{N\Delta}$ &
$\tilde{\beta}_{dh}^{NN}$ & \\
\hline Proton & $\frac{4}{3}D^2$ & $\frac{{\cal C}^2}{6}$ &
$-\frac{(3F-D)^2}{72m_N}M_0^2(\mu_D+3\mu_F)$ & \\
Neutron & -$\frac{4}{3}D^2$ & $-\frac{{\cal C}^2}{6}$ &
$\frac{(3F-D)^2}{36m_N}M_0^2\mu_D$ & \\
\hline
\noalign{\bigskip}
\hline
 & $\tilde{\beta}_{2\pi}^{NN}$ & $\tilde{\beta}_{2\eta}^{NN}$ &
$\tilde{\beta}_{2\pi}^{N\Delta}$ & $\tilde{\beta}_{h}^{NN}$ \\
\hline Proton &
$(\frac{31}{36}D^2-\frac{1}{4}F^2-\frac{1}{2}DF)\mu_D$ &
$\frac{3D^2+3F^2-2DF}{12}(\mu_D+3\mu_F)$ &
$\frac{5}{9}{\cal C}^2\mu_C$ & $\frac{(3F-D)(F-D)}{3}(\mu_D+3\mu_F)$ \\
Neutron & $-(\frac{11}{18}D^2-\frac{1}{2}F^2-\frac{19}{15}DF)\mu_D$
& $\frac{-3D^2-3F^2+2DF}{2}\mu_D$ &
$-\frac{5{\cal C}^2}{18}\mu_C$ & $\frac{2(3F-D)(D-F)}{3}\mu_D$ \\
\hline
\end{tabular}
\end{center}
\end{table}

The loop contribution to the nucleon magnetic moments at leading
order in the quenched case is expressed as
\begin{equation}
\tilde{G}_{M}^{p(\text
{LO})}=\frac{m_N}{8\pi^3f_\pi^2}\left[\tilde{\beta}_{1\pi(p)}^{NN}I_{1\pi}^{NN}
+\tilde{\beta}_{1\pi(p)}^{N\Delta}I_{1\pi}^{N\Delta}+\tilde{\beta}_{dh(p)}^{NN}I_{6\pi}^{NN}\right],
\end{equation}
\begin{equation}
\tilde{G}_{M}^{n(\text
{LO})}=\frac{m_N}{8\pi^3f_\pi^2}\left[\tilde{\beta}_{1\pi(n)}^{NN}I_{1\pi}^{NN}
+\tilde{\beta}_{1\pi(n)}^{N\Delta}I_{1\pi}^{N\Delta}+\tilde{\beta}_{dh(n)}^{NN}I_{6\pi}^{NN}\right],
\end{equation}
where
\begin{equation}
I_{6j}^{NN}=\int d^3k\frac{k^2u^2(\overrightarrow{k})}
{\omega_j^5(\overrightarrow{k})}.
\end{equation}
The NLO contribution can be written as
\begin{equation}
\tilde{G}_{M}^{p(\text
{NLO})}=\frac{m_N}{48\pi^3f_\pi^2}\left[\tilde{\beta}_{2\pi(p)}^{NN}I_{2\pi}^{NN}+\tilde{\beta}_{2\eta(p)}^{NN}I_{2\eta}^{NN}
+\tilde{\beta}_{2\pi(p)}^{N\Delta}I_{2\pi}^{N\Delta}+\tilde{\beta}_{h(p)}^{NN}I_{2\pi}^{NN}\right],
\end{equation}
\begin{equation}
\tilde{G}_{M}^{n(\text
{NLO})}=\frac{m_N}{48\pi^3f_\pi^2}\left[\tilde{\beta}_{2\pi(n)}^{NN}I_{2\pi}^{NN}+\tilde{\beta}_{2\eta(n)}^{NN}I_{2\eta}^{NN}
+\tilde{\beta}_{2\pi(n)}^{N\Delta}I_{2\pi}^{N\Delta}+\tilde{\beta}_{h(n)}^{NN}I_{2\pi}^{NN}\right].
\end{equation}
In the quenched case, the wave function renormalization constant is
obtained as $\tilde{Z}$
\begin{equation}
\tilde{Z}=1-\frac{1}{48\pi^3f_\pi^2}\left[\tilde{\beta}_{\pi}^{NN}I_{2j}^{NN}-\tilde{\beta}_{\pi}^{N\Delta}I_{2j}^{N\Delta}
-\tilde{\beta}_{dh}^{NN}I_{6j}^{NN}-\tilde{\beta}_{h}^{NN}I_{2j}^{NN}\right],
\end{equation}
where the coefficients $\tilde{\beta}$ are shown in table II. For the double hairpin diagram, $M_0$ is the
interaction strength.

Similar to the full QCD case, the quenched magnetic moments of the nucleon are expressed as
\begin{equation}\label{renorqu}
\tilde{\mu}_{p(n)} = \left\{a_0^{p(n)}+a_2^{p(n)}m_\pi^2+a_4^{p(n)}m_\pi^4 + \tilde{\mu}_{l1}^{p(n)}\right\}
/(2-\tilde{Z}-\frac{\tilde{\mu}_{l2}^{p(n)}}{\tilde{\mu}_{p(n)}^{\text{tree}}}),
\end{equation}
where $\tilde{\mu}_{l1}^{p(n)}$ is the loop contribution from diagrams (a) and (b) in Fig.~1 with quenched coefficients, while $\tilde{\mu}_{l2}^{p(n)}$
is the contribution from the other diagrams. Because the simulation is
on a lattice with length $L$ in the spatial dimensions, the momentum
integral is replaced by a discrete sum over the momentum, i.e.,
\begin{equation}
\int d^3k \Rightarrow \left(\frac{2\pi}{aL}\right)^3
\sum_{k_x,k_y,k_z},
\end{equation}
where the momenta $k_x$,$k_y$ and $k_z$ are given by $2\pi n/L$ and
the infinite sum is regulated by the finite-range regulator.  By
fitting the quenched lattice data with Eq.~(\ref{renorqu}), one can
get the parameters $a_i$. The full QCD results are then obtained with
Eq.~(\ref{renor}).

\section{Numerical results}

In the numerical calculations, the parameters are chosen as $D=0.76$
and $F=0.50$ ($g_A=D+F=1.26$). The coupling constant ${\cal C}$ is
chosen to be $-1.2$ which is the same as Ref.\ \cite{Jenkins2}.  The
regulator,  $u(k)$, may be chosen as a monopole, dipole or
Gaussian function, since all have been shown to yield
similar results~\cite{Young2}.  In our
calculations the dipole function is used:
\begin{equation}
u(k)=\frac1{(1+k^2/\Lambda^2)^2},
\end{equation}
with $\Lambda = 0.8$ GeV.

The $K$- and $\eta$-meson masses have relationships with the pion
mass according to
\begin{equation}
m_K^2=\frac12 m_\pi^2+m_K^2|_{\rm phy}-\frac12 m_\pi^2|_{\rm phy},
\end{equation}
\begin{equation}
m_\eta^2=\frac13 m_\pi^2+m_\eta^2|_{\rm phy}-\frac13 m_\pi^2|_{\rm
  phy}\, ,
\end{equation}
and enable a direct relationship between the meson dressings of the
nucleon magnetic moments and the pion mass.

\begin{figure}[tbp]
\begin{center}
\includegraphics[scale=0.8]{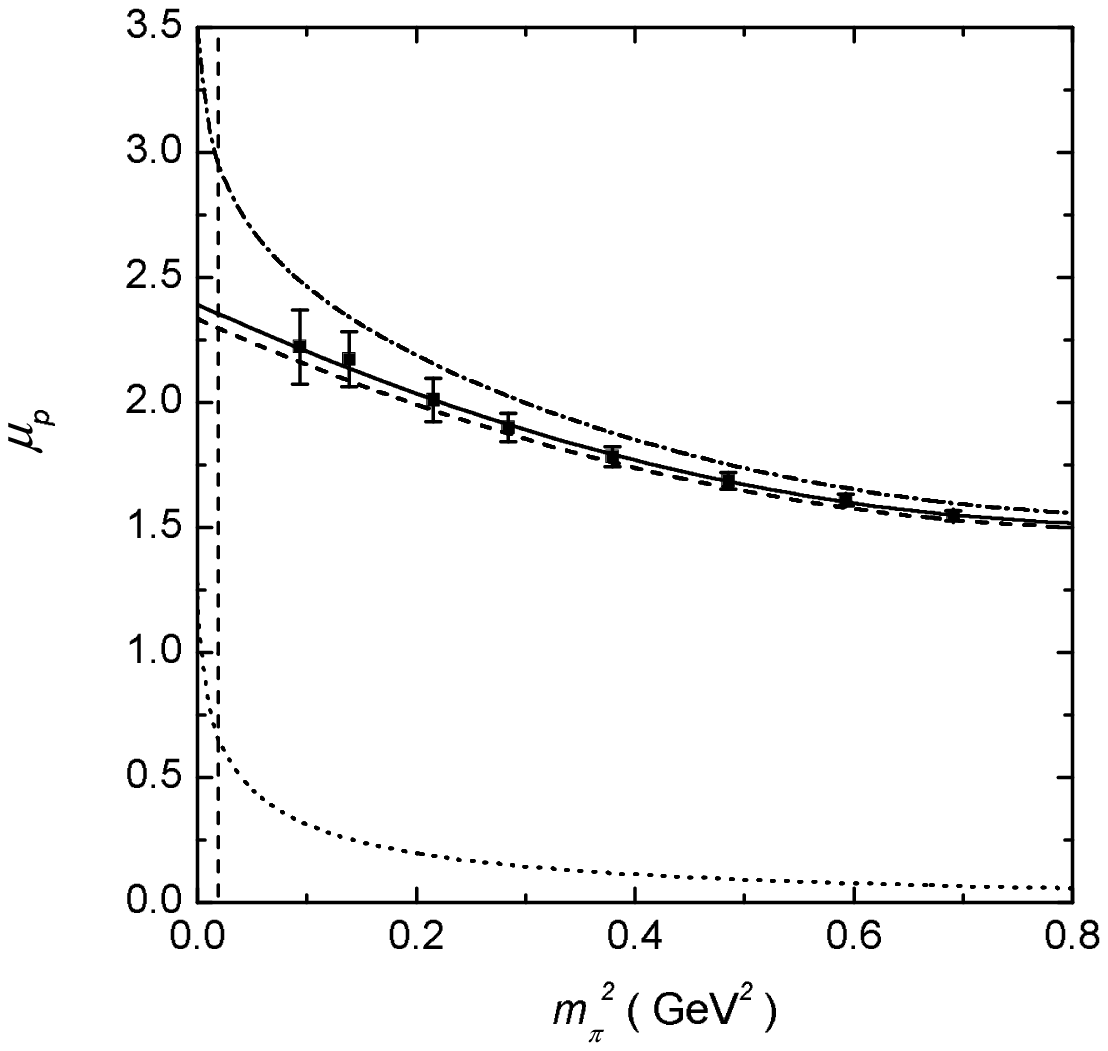}
\caption{The proton magnetic moment as a function of the squared pion
  mass.  The solid line illustrates the finite-volume quenched-QCD fit
  to the lattice results. The dashed, dotted and dash-dotted lines
  correspond to the infinite-volume full-QCD results at tree level,
  leading loop and sum of tree level and leading loop, respectively.}
\end{center}
\end{figure}

We begin by considering nucleon magnetic moments from the CSSM
Lattice Collaboration \cite{Boinepalli:2006xd}.  The leading order
result of the proton magnetic moment versus $m_\pi^2$ is shown in
Fig.~2. The solid line is for the finite-volume quenched-QCD fit and the dashed,
dotted and dash-dotted lines are for the infinite-volume full QCD results of tree
level, leading loop and sum of tree level and leading loop,
respectively. One can see that quenched lattice results can be described
very well in quenched chiral effective field theory.  At the physical pion
mass, the proton magnetic moment $\tilde{\mu}_p$ is about $2.25$
which is significantly smaller than the experimental data. With the
obtained fitting parameters $a_i$, the full QCD results are determined
and illustrated in the figure.

In the quenched case, the loop contribution is small.  While
quenched-QCD coefficients of nonanalytic terms are typically smaller
than in the full QCD case, the dominant effect here is that the
momentum integration is replaced by the finite-volume sum.  The loop
contribution in full QCD gives the dominant curvature of the pion mass
dependence of the proton magnetic moment.  The proton magnetic moment in
the full QCD case is significantly larger.  At the physical pion mass,
the proton magnetic moment in the full QCD case, $\mu_p$, is
approximately $2.95\ \mu_N$ which is similar to the experimental
value, $2.79\ \mu_N$.

\begin{figure}[tbp]
\begin{center}
\includegraphics[scale=0.8]{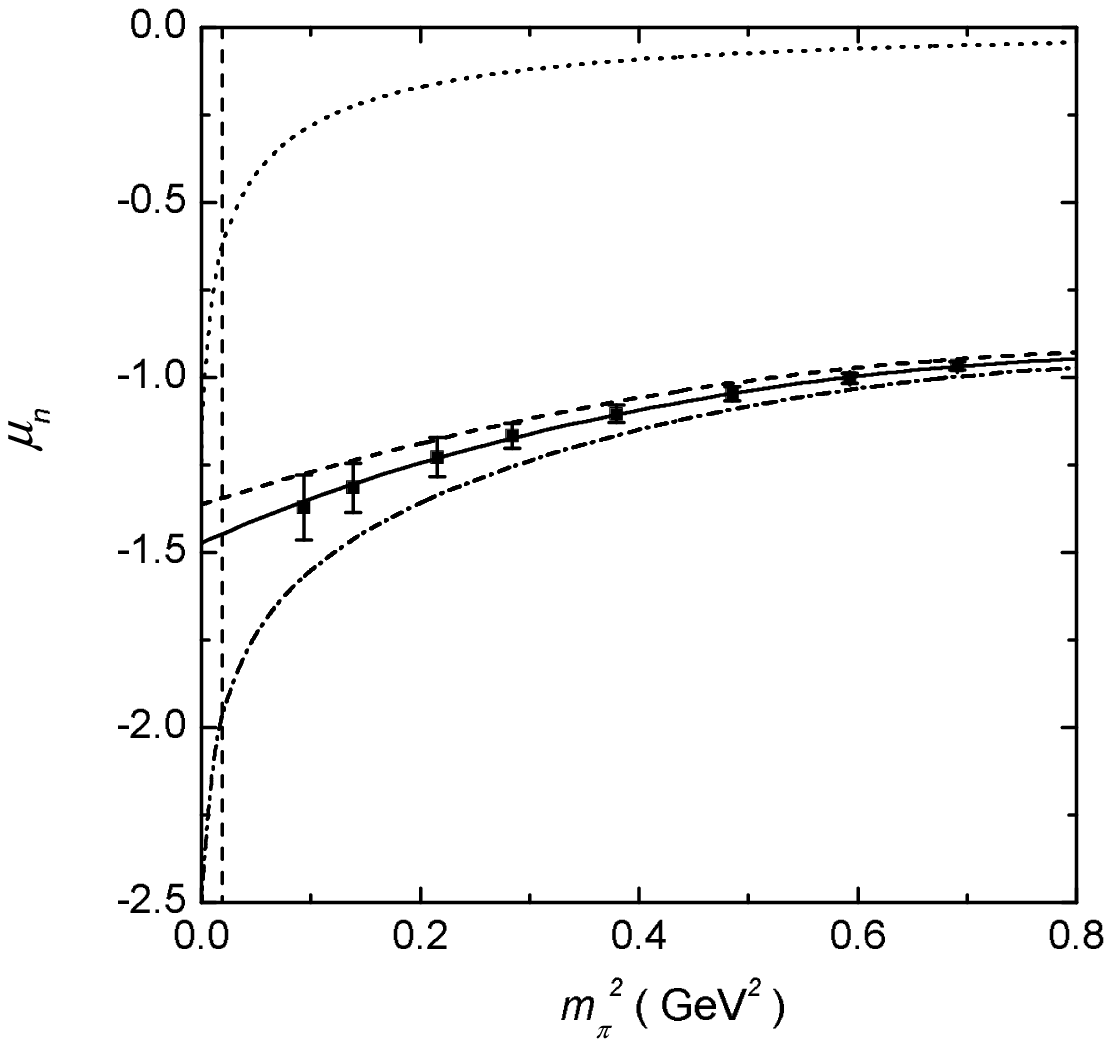}
\caption{The neutron magnetic moment as a function of the squared pion
  mass.  The solid line illustrates the finite-volume quenched-QCD
  fit. The dashed, dotted and dash-dotted lines correspond to the
  infinite-volume full-QCD results of tree level, leading loop and sum
  of tree level and leading loop contribution, respectively.}
\end{center}
\end{figure}

The leading order result for the neutron magnetic moment versus $m_\pi^2$
is shown in Fig.~3.  Again, the finite-volume quenched-QCD lattice
results are described very well by finite-volume finite-range
regularised quenched chiral effective field theory.  The curvature of
the line is small.  At the physical pion mass, the finite-volume
quenched neutron magnetic moment is around $-1.5$.  The associated
full QCD results of tree level, leading loop and sum of tree level and
leading loop are shown as well.  Similar to the proton case, the loop
contribution changes smoothly at large pion mass and drops quickly at
small pion masses. The total value of neutron magnetic moment from our
leading-order calculations is $\mu_n \simeq -1.96\ \mu_N$, similar to
the physical value of $-1.91\ \mu_N$.

\begin{figure}[tbp]
\begin{center}
\includegraphics[scale=0.8]{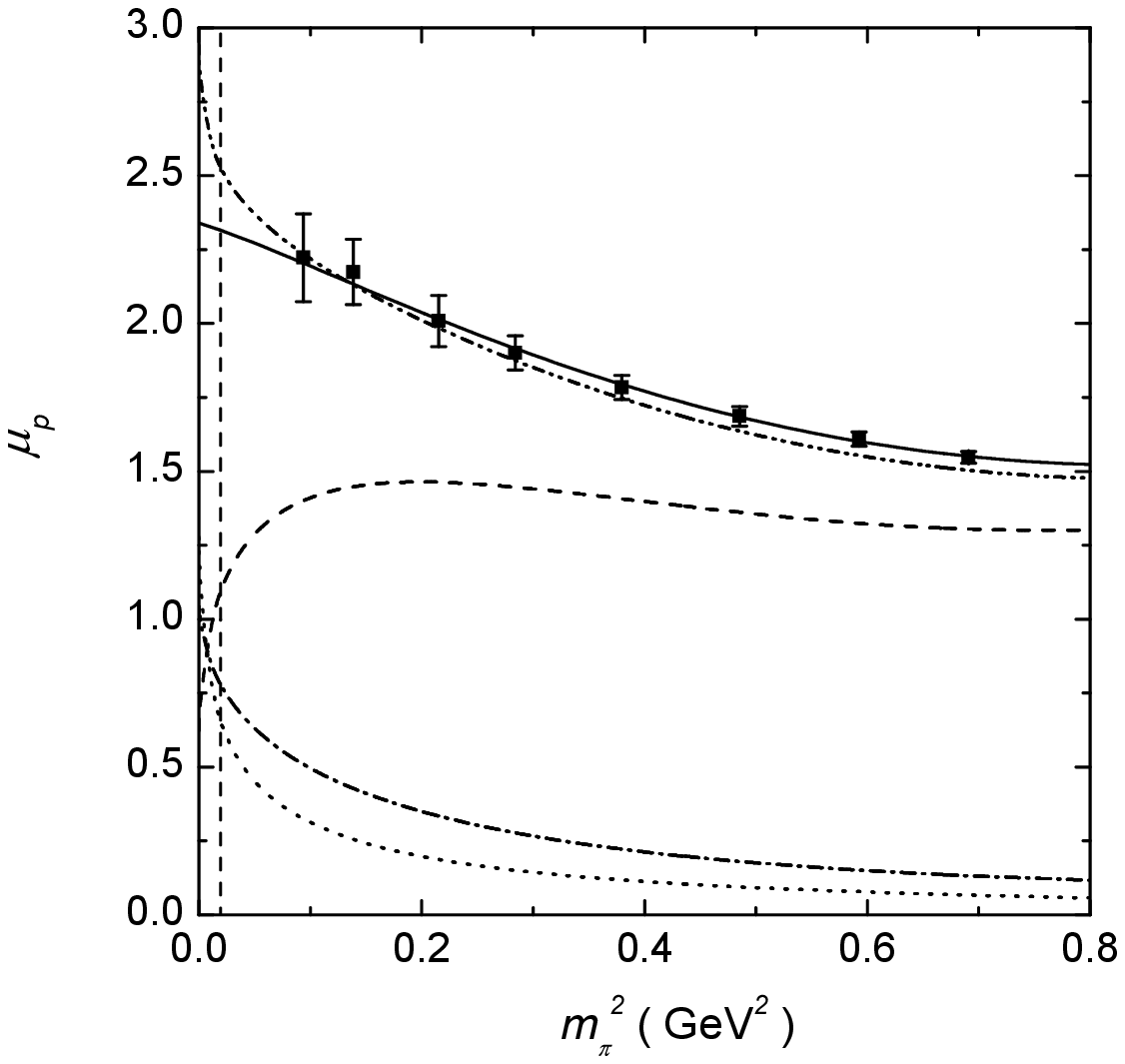}
\caption{The proton magnetic moment versus squared pion mass.  The
  solid line illustrates the finite-volume quenched-QCD fit. The
  dashed, dotted, dash-dotted and dash-dot-dotted lines correspond to
  the infinite-volume full-QCD results of tree level, leading loop,
  NLO loop and sum of tree level and loop contribution,
  respectively.}
\end{center}
\end{figure}

The NLO result for the proton magnetic moment versus
$m_\pi^2$ is shown in Fig.~4. The solid line is the finite-volume
quenched-QCD result.  The dashed, dotted, dash-dotted and
dash-dot-dotted lines are for the infinite-volume full-QCD results of
tree level, leading order loop, NLO loop and sum of
tree and loop contribution, respectively. At NLO, the
quenched lattice results continue to be described well by
finite-volume quenched chiral effective field theory.  However, at
NLO, the approach to the chiral limit displays some
downward curvature associated with the new wave-function
renormalization contributions which appear only at NLO.
The wave function renormalization constant $Z$ decreases
quickly at small pion mass.

At the physical pion mass, the infinite volume tree-level
contribution to the proton magnetic moment changes from $2.30$ to
$1.10$. The leading loop contribution at the physical pion mass is
$0.65$. The NLO loop contribution has a smaller curvature
than the leading loop. It contributes $0.78$ to the proton magnetic
moment. The sum of tree level, leading loop and NLO loop
contribution to the proton magnetic moment is $2.53\ \mu_N$ to be
compared with the experimental value of $2.79\ \mu_N$.

\begin{figure}[tbp]
\begin{center}
\includegraphics[scale=0.8]{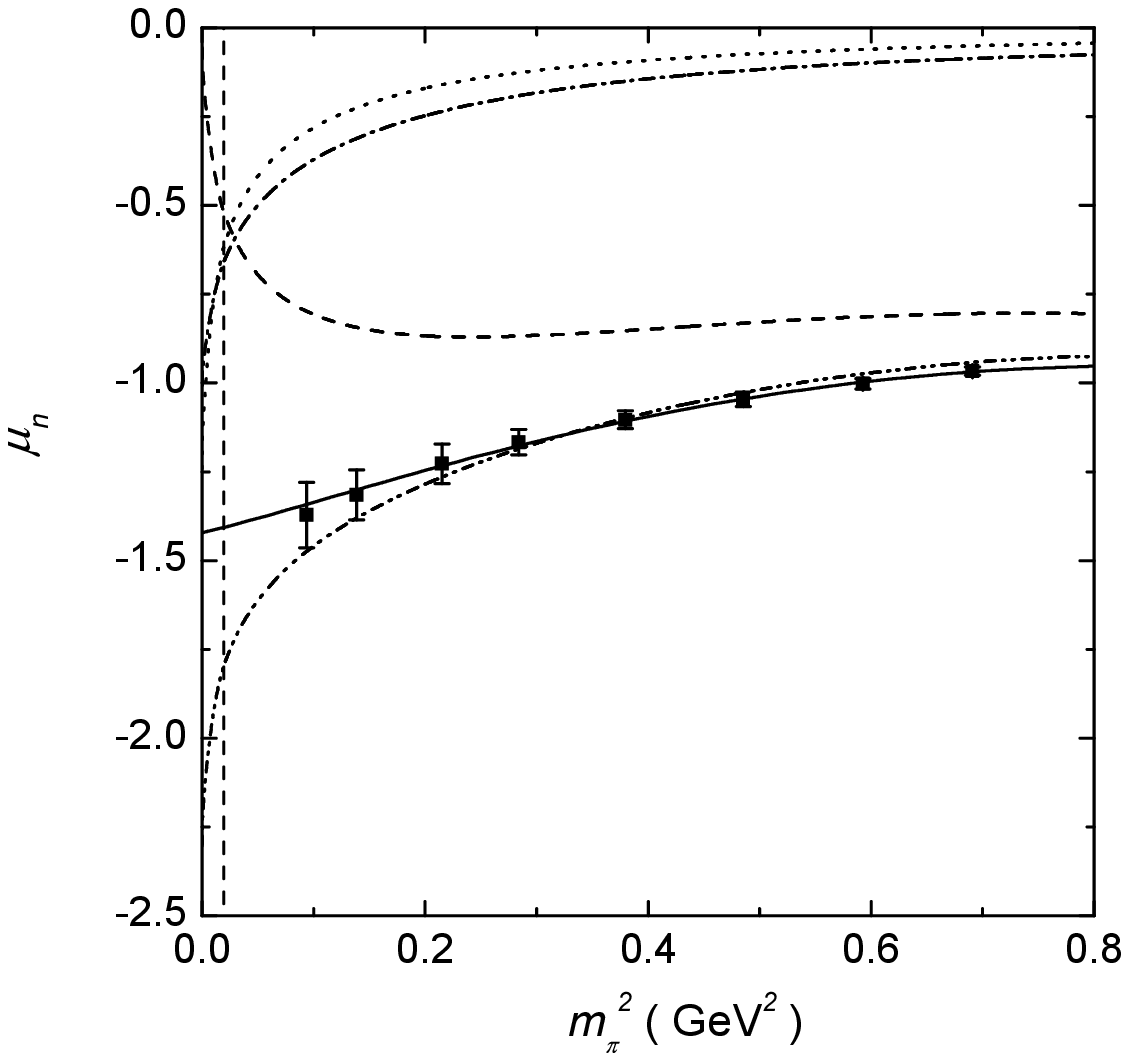}
\caption{The neutron magnetic moment versus squared pion mass.  The
  solid line illustrates the quenched-QCD fit. The dashed, dotted,
  dash-dotted and dash-dot-dotted lines correspond to the
  infinite-volume full-QCD results of tree level, leading loop,
  NLO loop and sum of tree level and loop contribution,
  respectively.}
\end{center}
\end{figure}

The NLO result for neutron magnetic moment $\mu_n$
versus $m_\pi^2$ is shown in Fig.~5.  The meaning of the different
types of lines are the same as for Fig.~4.  Here the wave-function
renormalization has a more subtle effect.  As anticipated, the
NLO loop contribution has a smaller curvature than the
leading-order loop contribution.  At the physical pion mass, the tree
level, leading loop and NLO loop contribute to the neutron
magnetic moment $-0.52$, $-0.62$ and $-0.66\ \mu_N$, respectively.
The total neutron magnetic moment at NLO is
$-1.80\mu_N$ which remains close to the experimental value of
$-1.91\ \mu_N$.

We should mention that when we calculate the NLO loop
contribution the tadpole diagram is not included explicitly.
That is, the
tadpole contribution is handled by adjusting the
parameters $a_i'$ to $a_i$, so that:
\begin{equation}
\label{tad}
a_0'+a_2'm_\pi^2+a_4'm_\pi^4 + \mu_{tad} \simeq a_0+a_2m_\pi^2+a_4m_\pi^4 \, ,
\end{equation}
where $\mu_{tad}$ is the tadpole contribution to the magnetic moments.
In Ref.~\cite{Wang1}, the chiral extrapolation did explicitly include
this tadpole diagram but the numerical results were almost the same if we
refit the lattice data without this diagram. This means that,
in practice, the new
parameters, $a_i$, can compensate the contribution of the
tadpole diagram. In the present work we explored both the explicit
and implicit inclusion of the tadpole term,
with the numerical results clearly favoring
the approximation where the fitting parameters on the right hand side of
Eq.~(\ref{tad}) are the same in the quenched and full QCD cases.

Chiral symmetry can be realized in a number of ways, resulting
in different forms for the effective Lagrangian.
In Ref.~\cite{Jennings}, the authors applied two different
Lagrangian densities incorporating chiral symmetry to the
problem of pion-nucleon scattering, with the nucleon represented
by an MIT bag.
In one case the interaction was confined to the bag surface,
where only a Yukawa
type $NN\pi$ interaction appeared. The other formulation involved a
volume-interaction, where a contact term (four particle $NN\pi\pi$
term) is required.  Their conclusion was that transforming from the surface
interaction to the volume interaction amounts to summing the
contribution from all excited intermediate states of the confined quarks.
That is, the two formulations give equivalent results if excited
intermediate states are included.  One can also study the magnetic
moments with the pseudoscalar nucleon-meson interaction where no
tadpole diagram appears -- c.f. Refs.~\cite{Lu,Lyubovitskij,Faessler}.
With this background we conclude that the tadpole contribution to the
magnetic moments from the contact term corresponds to the contribution
of diagram (c) in Fig.~1 summed over an infinite set of highly excited
baryon states and phenomenologically this appears to be appropriately
incorporated through Eq.~(\ref{tad}).

\section{Summary}

\begin{table}
\caption{The obtained coefficients $a_i$  and magnetic moments.}
\begin{center}
\begin{tabular}{||c|c|c|c|c|c|c|c|c||}
\hline
& $a_0$ & $a_2$ & $a_4$ & $\tilde{\mu}^{\text {LO}}$ & $\tilde{\mu}^{\text {NLO}}$ & $\mu^{\text {LO}}$ & $\mu^{\text {NLO}}$ & Exp. \\
\hline
Proton & $2.34$ & $-2.08$ & $1.24$ & 2.35 & 2.31 & 2.95 & 2.53 & 2.79 \\
Neutron & $-1.39$ & $1.17$ & $-0.70$ & $-1.45$ & $-1.41$ & $-1.96$ & $-1.80$ & $-1.91$ \\
\hline
\end{tabular}
\end{center}
\end{table}

We have extrapolated quenched lattice QCD results for nucleon magnetic
moments extending into the chiral regime~\cite{Boinepalli:2006xd} to
the physical pion mass using finite-volume finite-range regularised
chiral effective field theory.  Here, the NLO
contributions are included, with the numerical
results showing that the quenched lattice results are described very well.
By fitting quenched lattice data, the parameters $a_i$ can be obtained
and using the dipole regulator parameter of 0.8 GeV the full QCD
results are predicted.  The infinite-volume full QCD results obtained
at the physical pion mass are in reasonable agreement with
experiment at both leading order and NLO.
Thus finite-range regularised chiral effective field theory
provides an effective formalism for extending chiral perturbation
theory beyond the power-counting regime and connecting quenched QCD
and full QCD in a quantitative manner.
The parameters and results are summarised in Table IV.

It is interesting how the NLO contributions come in a
compensating fashion.  While each NLO contribution
displays significant curvature in the chiral regime, the net
contribution is relatively smooth and otherwise easily compensated for
by the residual series expansion.  We expect that this
qualitative behavior will continue as additional higher-order terms
are introduced, as we are informed by the lattice QCD results
displaying a smooth slowly-varying quark mass dependence.  Indeed, it
will be interesting to examine more physical quantities to gain a
deeper understanding of the utility of finite-range regularised chiral
effective field theory.

\section*{Acknowledgments}

This work is supported in part by DFG and NSFC (CRC 110), by the
National Natural Science Foundation of China (Grant No. 11035006) and
by the Australian Research Council through grants FL0992247 (AWT),
DP110101265 (DBL and RDY) and through the ARC Centre of Excellence for
Particle Physics at the Terascale.

\end{document}